\documentclass[sigconf]{acmart}
\setcopyright{none}
\renewcommand\footnotetextcopyrightpermission[1]{}

\usepackage{booktabs}

\begin{document}

\title{A Global Author-Identity Map for the World of Code:\\
       62.7M Developer Identities from 106.8M Author Strings over 5.87B Commits}

\author{Audris Mockus}
\affiliation{%
  \institution{University of Tennessee, Knoxville}
  \city{Knoxville}\state{TN}\country{USA}}
\email{audris@utk.edu}

\begin{abstract}
Mining software repositories at global scale founders on author identity: the
same developer commits under many name/email strings, and the same string is
reused by many developers. We release a curated author-identity map for the
World of Code (WoC) version \texttt{V2604}, covering all
$5{,}866{,}595{,}698$ commits in the collection. The release has four
co-versioned artifacts: (i) a global alias map, \texttt{a2AFullSUG}, that
folds $106{,}826{,}059$ raw author/committer strings into canonical
identities; (ii) a per-identity classification, \texttt{A2clsFull}, tagging
each id as \emph{good}, \emph{bad-by-attribute}, \emph{local}, \emph{bot}, or
\emph{partial}; (iii) a within-project resolution table, \texttt{P2aAFull},
that recovers low-quality ids inside the single project where their reuse is
unambiguous; and (iv) a commit-to-identity table, \texttt{c2AFull}, that tags
every commit with the provenance of its resolution. The map is
\emph{mega-cluster free}; its largest cluster is $6{,}910$ ids, a single
GitHub \texttt{noreply} identity. It resolves $73.5\%$ of all six billion
commits into a multi-id identity, raising human-id commit coverage to
$98.17\%$. The central design problem is \emph{clumping} rather than recall: the
naive transitive union over shared-attribute edges welds three million
unrelated people into one cluster, an over-merge that recall-only benchmarks
price at zero. We therefore report \emph{both} error families (splitting and
clumping) and show that the high precision claimed by global-scale union maps
can be an artifact of never measuring the conflated
region. Against the ALFAA human-rated gold set the released map scores recall
$0.70$ / precision $0.88$ with a largest cluster of $6{,}910$, where the prior
WoC map's apparent precision of $0.95$ collapses to $0.52$ once its
$3{,}006{,}318$-id mega-cluster is counted. Finally, a canonical
software-author identity is a join key \emph{across} corpora: we discuss linking
these $6.3\times10^7$ canonical developer identities (folded from $10^8$ raw ids)
to scholarly author graphs ($1.1\times10^8$
Semantic Scholar and the OpenAlex authors), a setting where clumping, not
recall, is again the binding constraint. All artifacts ship with the WoC
\texttt{V2604} release and a self-contained replication package.
\end{abstract}


\maketitle

% ===========================================================================
\section{Introduction}
\label{sec:intro}
Authorship is the join key of empirical software engineering: contribution
counts, bus-factor estimates, developer-network studies, and provenance
analyses all key on \emph{who} authored a commit. In a single repository the
author string is a serviceable key. Across the whole of public version
control it is not. A developer signs commits as
\texttt{Jane Doe <jane@work.com>}, \texttt{jdoe <jane@personal.org>}, and
\texttt{Jane <jane@users.\allowbreak noreply.\allowbreak github.com>}; conversely
a thousand strangers all commit as \texttt{root@localhost} or
\texttt{Your Name <you@\allowbreak example.com>}.
Resolving these (merging the aliases of one person without welding strangers
together) is the \emph{author de-aliasing} problem, and it grows qualitatively
harder with scale: at $10^8$ identities a single careless transitive merge can
fuse millions of unrelated people into one ``mega-cluster.''

The World of Code (WoC)~\cite{ma2021world} is a periodically updated census of
public version control, deduplicated at the git-object level. Its
\texttt{V2604} release records $5{,}866{,}595{,}698$ commits authored under
$106{,}826{,}059$ distinct author/committer strings. We release a curated
\emph{author-identity map} over this collection and the supporting tables a
consumer needs to use it correctly: a per-identity quality classification, a
within-project resolution layer for the ids the global map cannot safely merge,
and a commit-level table that records, for every one of the six billion
commits, how its author was resolved.

This paper is a data description. Section~\ref{sec:dataset} specifies the four
artifacts: their schemas, sizes, and access. Section~\ref{sec:construction}
describes the production pipeline that builds them. Section~\ref{sec:rationale}
is the paper's argument: every non-obvious choice in that pipeline was forced
by a measured failure of the simpler alternative, and we summarize the
eighteen-experiment record (logged in full in the companion methodology
paper~\cite{aliasingmethod}) that establishes them. Section~\ref{sec:quality}
reports external validation, develops the splitting/clumping distinction, and
argues that the precision claimed by global-scale union maps is unmeasured
clumping. Section~\ref{sec:downstream} quantifies how the choice of map shifts
downstream analytics---developer counts, productivity, team resilience, and
centrality. Section~\ref{sec:impact} turns to broader impact: using the map as a
join key \emph{across} corpora to link software authors to their scholarly
publications. Sections~\ref{sec:usage}--\ref{sec:concl} give usage,
limitations, and conclusions.

\paragraph{What is new relative to prior WoC identity maps.} Earlier WoC
releases shipped alias maps built by string-similarity union (the ALFAA
lineage~\cite{amreen2019alfaa,fry2020dataset}). Those maps achieved high recall
but at the cost of large over-merged clusters that silently corrupt any
per-author aggregate (Section~\ref{sec:rationale}). The map released here is the
first WoC author map that is simultaneously (a) mega-cluster free by
construction, (b) shipped with a per-id quality classification and a
within-project fallback, and (c) validated on two independent ground truths
read jointly for precision \emph{and} recall.

\paragraph{Research questions.} Beyond releasing the artifacts, the paper answers
four questions about author identity at global scale:
\begin{description}
\item[RQ1] What does it take to build a global author map that is
  \emph{simultaneously} high-recall and free of over-merge? (Section~\ref{sec:rationale})
\item[RQ2] How should a disambiguation result at this scale be \emph{evaluated},
  given that the dominant benchmarks measure only one of the two error families?
  (Section~\ref{sec:quality})
\item[RQ3] How much, and \emph{in which direction}, does the choice of identity
  method change the downstream developer analytics that consume it, across
  realistic scenarios (head-count, productivity, bus factor, collaboration
  networks, centrality, reviewer/successor recommendation)?
  (Section~\ref{sec:downstream})
\item[RQ4] Can a calibrated, mega-cluster-free map serve as a join key
  \emph{across} corpora, linking software authors to the scholarly record, where
  clumping, not recall, is the binding constraint? (Section~\ref{sec:impact})
\end{description}
RQ3 is the question a practitioner faces when deciding whether aliasing is worth the
trouble; we answer it by comparing three method archetypes (under-merge, over-merge,
calibrated) on the same corpus and showing the first two err in opposite directions
on every task.

% ===========================================================================
\section{The Dataset}
\label{sec:dataset}
The release comprises four gzip-compressed, \texttt{;}-separated, \texttt{LC\_ALL=C}
sorted tables, all keyed to WoC version \texttt{V2604} and sharded for parallel
access. Table~\ref{tab:artifacts} summarizes them.

\begin{table}[t]
\centering\small
\caption{Released artifacts (WoC \texttt{V2604}).}
\label{tab:artifacts}
\begin{tabular}{@{}llr@{}}
\toprule
Artifact & Record (\texttt{;}-separated) & Rows / size \\
\midrule
\texttt{a2AFullSUG} & \texttt{rawId;canonicalId} & $106{,}826{,}059$ ids \\
\texttt{A2clsFull}  & \texttt{member;canonical;class} & $106{,}826{,}059$ ids \\
\texttt{P2aAFull}   & \texttt{project;rawId;A;rule} & $5{,}359{,}181$ asgmts \\
\texttt{c2AFull}    & \texttt{commit;A;prov} & $5{,}866{,}595{,}698$ cmts \\
\bottomrule
\end{tabular}
\end{table}

\paragraph{(1) Global alias map \texttt{a2AFullSUG}.} The core artifact (the
\texttt{SUG} suffix records its edge composition: shingle union plus ghid,
defined in Section~\ref{sec:construction}). Each
row maps a raw author/committer string to its canonical representative
\texttt{A} (the highest-quality member of its identity cluster, selected by a
deterministic rule favoring a real name and a non-generic email). An id that is
its own representative maps to itself. The map induces an equivalence relation:
two raw ids are the same person iff they share a canonical \texttt{A}. The map
is mega-cluster free (largest cluster $6{,}910$ ids, itself a single GitHub
\texttt{noreply} identity, and the $>$$10$k size bin is empty), so per-author
aggregates computed over it are not silently contaminated by an over-merge.

\paragraph{(2) Per-identity classification \texttt{A2clsFull}.} For every id,
its canonical representative and a quality class under a five-way taxonomy
(Table~\ref{tab:classes}). The class tells a consumer \emph{why} an id is or is
not a global representative: \emph{good} ids carry a real name and email;
\emph{bad-by-attribute} ids carry a generic name or a shared/placeholder
attribute; \emph{local} ids carry a machine-local address with a real
username; \emph{bot} ids are automated accounts; \emph{partial} ids are
otherwise-good but carry only one of name/email. The three non-good, non-bot
classes are the $5.46$M ids the global map declines to merge across
projects and instead maps to self: the pool the within-project layer targets.

\begin{table}[t]
\centering\small
\caption{Per-id quality classes (\texttt{A2clsFull}).}
\label{tab:classes}
\begin{tabular}{@{}lrr@{}}
\toprule
Class & Ids & Share \\
\midrule
good            & $100{,}814{,}372$ & $94.37\%$ \\
bad-by-attribute & $2{,}652{,}369$  & $2.48\%$ \\
local           & $2{,}562{,}118$   & $2.40\%$ \\
bot             & $553{,}736$       & $0.52\%$ \\
partial         & $243{,}464$       & $0.23\%$ \\
\midrule
total           & $106{,}826{,}059$ & $100\%$ \\
\bottomrule
\end{tabular}
\end{table}

\paragraph{(3) Within-project resolution \texttt{P2aAFull}.} An id that is
ambiguous \emph{across} projects (\texttt{root}, \texttt{ubuntu},
\texttt{user@laptop}) is usually unambiguous \emph{within} the single deforked
project where it appears, because reuse there is local. This table assigns such
an id to a canonical person inside one project, with the \texttt{rule} field
recording how: \emph{anchor} (a unique lowercased full-name or email-local-part
match to exactly one already-resolved person in the project) or \emph{solo}
(the project has a single committing person). It holds $5{,}359{,}181$
assignments ($2{,}355{,}347$ anchor $+$ $3{,}003{,}834$ solo), recovering
$47.8\%$ of the low-quality commit pool the global map leaves unresolved.

\begin{figure}[t]
\centering
\includegraphics[width=0.92\linewidth]{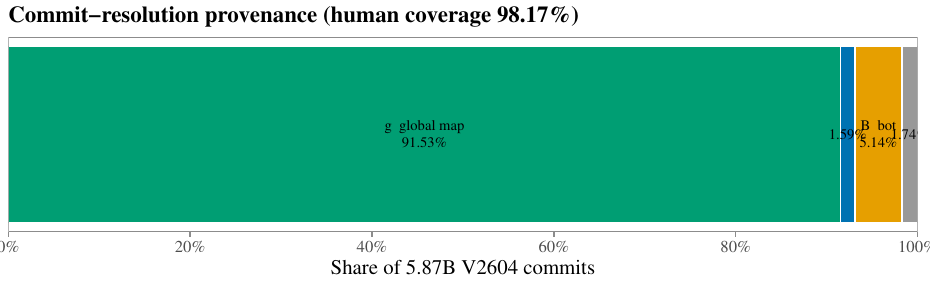}
\caption{How every one of the $5{,}866{,}595{,}698$ \texttt{V2604} commits is
resolved, by provenance tag. The global map carries the bulk; the within-project
layer adds the harder ids it cannot safely merge, lifting human-id coverage to
$98.17\%$.}
\label{fig:provenance}
\end{figure}

\paragraph{(4) Commit-to-identity table \texttt{c2AFull}.} The consumer-facing
join: for every one of the $5{,}866{,}595{,}698$ \texttt{V2604} commits, its
resolved author \texttt{A} and a one-character provenance tag recording how the
resolution was obtained: \texttt{g} global map, \texttt{B} bot, \texttt{p}
within-project, \texttt{s} self/unresolved. The provenance distribution is
$g=91.53\%$, $B=5.14\%$, $p=1.59\%$, $s=1.74\%$. The tag lets a study filter to
the resolution quality it needs (e.g.\ drop \texttt{B}, keep
\texttt{g}$+$\texttt{p}, treat \texttt{s} as unknown) without re-deriving
anything (Figure~\ref{fig:provenance}).

\paragraph{Access and format.} All tables are part of the WoC \texttt{V2604}
release and follow WoC conventions: gzip, \texttt{;}-separated,
\texttt{LC\_ALL=C} sorted on the first field, section-sharded $0..127$ (maps)
or $0..31$ (derived) by a stable hash of the key for parallel scans. A
self-contained replication package (Section~\ref{sec:construction}) accompanies
the release.

% ===========================================================================
\section{Construction}
\label{sec:construction}
\begin{figure}[t]
\centering
\includegraphics[width=0.82\linewidth]{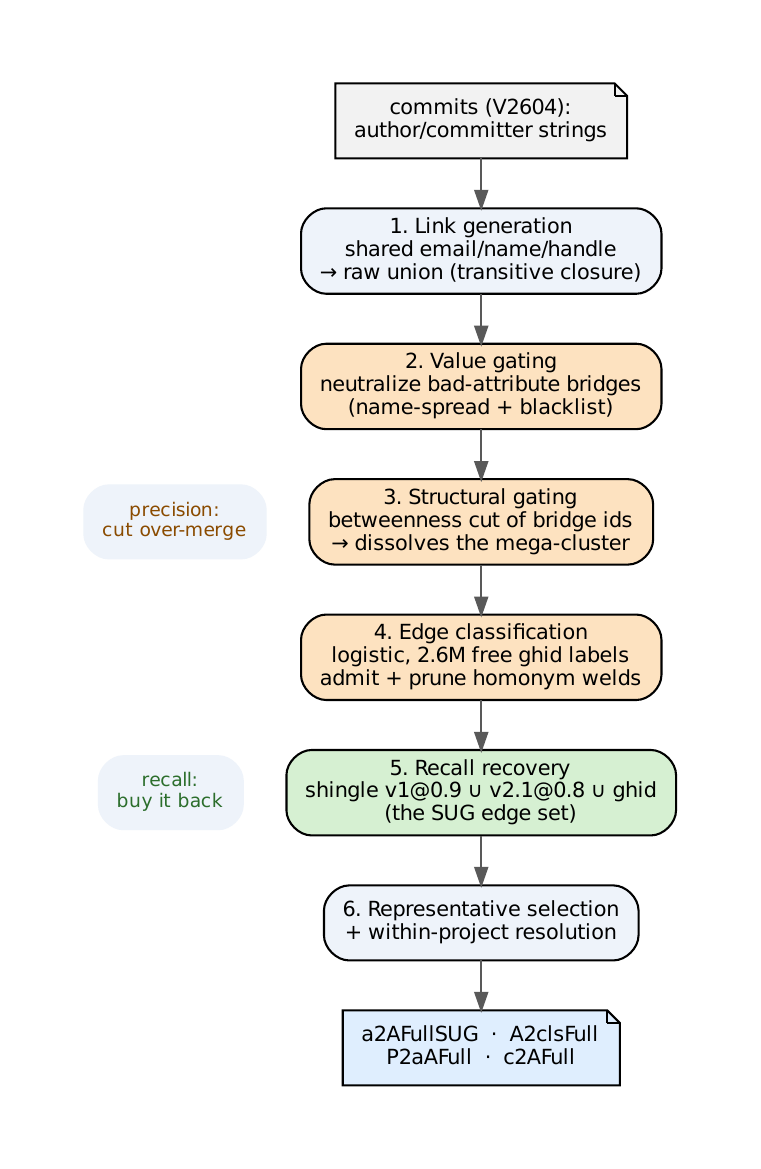}
\caption{The six-stage construction. Value, structural, and edge gating (orange)
attack over-merge; recall recovery (green) then admits the shingle and
\texttt{ghid} edges that the precision work held out, yielding the released
\texttt{a2AFullSUG} map and its companions.}
\label{fig:construction}
\end{figure}

The map is built by a six-stage pipeline over the WoC \texttt{V2604}
commit data (Figure~\ref{fig:construction}). We state the pipeline here as the
data description; Section~\ref{sec:rationale} justifies each stage.

\begin{enumerate}
  \item \textbf{Link generation.} From every commit, emit candidate
    same-person edges between author/committer strings that share evidence
    (exact email, exact name, or a shared rare handle). The transitive closure
    of these edges is the raw union.
  \item \textbf{Value gating.} Neutralize ids whose \emph{attributes} cannot
    identify a person (generic names, placeholder and shared emails, bot
    accounts) so they do not act as merge bridges. Detection is by name-spread
    and a frequency/blacklist battery.
  \item \textbf{Structural gating.} On the exact union graph, compute
    sampled betweenness and cut the small set of high-betweenness
    ``bridge'' ids that weld otherwise-disjoint communities. This dissolves the
    over-merged mega-cluster that value gating alone cannot.
  \item \textbf{Edge classification.} Score every candidate within-group pair
    with a logistic edge classifier (features: shared-attribute spread,
    name$\leftrightarrow$handle patterns, given-name-homonym flag, deforked
    project overlap), trained on $2.6$M free within-handle GitHub-id labels.
    Admit confident pairs; prune residual homonym welds.
  \item \textbf{Recall recovery.} Expand the dormant cross-project
    \emph{shingle} layer (candidate same-person links proposed by shared
    name/email token shingles but held out of the base union) into member
    pairs, score each with the classifier, and admit confident pairs as
    \emph{new} union edges; union with GitHub's own \texttt{noreply}
    same-account assertions. Two shingle-expansion variants contribute on
    disjoint margins and compose (Exp.~17), each admitted at a separately tuned
    threshold: the pairwise expansion \texttt{v1} at $\tau{=}0.9$ (the largest
    single recall gain in the record) and the map-level expansion \texttt{v2.1}
    at $\tau{=}0.8$ (at $0.9$ the map-level variant sacrifices GitHub recall).
    This is the production edge set: \texttt{v1@0.9} $\cup$ \texttt{v2.1@0.8}
    $\cup$ \texttt{ghid}, whose composition is what the map's \texttt{SUG}
    suffix records (shingle union $+$ ghid).
  \item \textbf{Representative selection and classification.} Choose a
    canonical representative per cluster by a quality rule (streaming, to bound
    memory); emit \texttt{a2AFullSUG} and \texttt{A2clsFull}. Resolve the
    self-mapped low-quality pool inside individual projects
    (\texttt{P2aAFull}); tag every commit with provenance (\texttt{c2AFull}).
\end{enumerate}

The pipeline runs as WoC SLURM jobs over the section-sharded commit tables. The
replication package ships the complete script set: link generation and gating,
bad-attribute detection, the C++ structural toolchain (betweenness/bridge cut),
the classifier training and scoring, the union and map builders, the
within-project matcher, and the evaluation harness.

% ===========================================================================
% DEFORKING CARVED OUT (2026-06-23): the project-graph deforking construction +
% external validation (Exps D1-D5) is now its own MSR Data & Tool Showcase at
% ~/papers/DeforkingShowcase/ (main.tex \input{deforking}). This data-track
% paper reverts to PURE author-identity. To re-fold, uncomment the line below.
% \input{deforking}

% ===========================================================================
\section{Design Rationale}
\label{sec:rationale}
Every non-default choice above replaced a simpler alternative that we measured
to fail. The full record is eighteen experiments~\cite{aliasingmethod};
Table~\ref{tab:decisions} maps each production decision to the experiment that
forced it. Two findings frame the rest: at $10^8$ ids de-aliasing is
\emph{two} problems, precision (do not over-merge) and recall (do not miss
aliases), and they must be solved in that order, because a recall mechanism
applied to an over-merged graph only makes the mega-cluster worse.

\begin{table}[t]
\centering\small
\caption{Production decisions and the experiment that justifies each
(full log in~\cite{aliasingmethod}).}
\label{tab:decisions}
\begin{tabular}{@{}p{0.46\columnwidth}p{0.46\columnwidth}@{}}
\toprule
Production decision & Justifying finding \\
\midrule
Gate \emph{before} unioning &
Ungated union welds a $3$M-id mega-cluster. \\
Do not gate on information score &
Score cutoffs destroy recall (Exp.~2). \\
Value gating is necessary but \emph{insufficient} &
Spread/degree gates slow-peel; neutralizing every bad attribute still leaves
the mega $94\%$ intact (Exps.~3--4, 8--10). \\
Add a \emph{structural} cut &
Sampled betweenness finds $\sim$$2$k bridge ids, $98\%$ attribute-clean,
and dissolves the mega (Exp.~11). \\
Resolve residue with an \emph{edge} classifier, not more value rules &
A classifier on $2.6$M GitHub-id labels prunes the homonym fragments the cut
isolates (Exps.~6, 12). \\
Recover recall by classifier-filtered shingle expansion at $\tau{=}0.9$ &
Largest single recall gain in the record; precision rises (Exp.~14). \\
\emph{Re-check} structure after adding edges &
At $\tau{=}0.5$ the mega returns: a cut certifies the edge set it saw, not
the design (Exp.~14). \\
Compose multiple edge sources &
v1, v2.1, and ghid gains live on disjoint margins and compose (Exp.~17). \\
Add a within-project layer &
Recovers $47.8\%$ of the low-quality pool the global map declines (Exp.~18). \\
Grade on \emph{two} benchmarks jointly &
A recall-only GT prices a $3$M mega at zero and ranks maps in reverse of a
precision audit (Exps.~13, 16). \\
\bottomrule
\end{tabular}
\end{table}

\paragraph{The over-merge phenomenon.} The naive transitive union over all
shared-attribute edges produces a single cluster of roughly three million
identities: real people fused by a mesh of generic strings. Any per-author
metric computed over such a map is meaningless for everyone in the cluster, and
the cluster is invisible to recall-only evaluation. Removing it is the
precision problem, and it resisted every mechanism that acts on \emph{nodes} or
\emph{values}: information-score cutoffs collapsed recall (Exp.~2);
project-spread and link-degree gates preserved recall but slow-peeled, each
threshold merely exposing the next moderate bridge (Exps.~3--4); and exhaustive
neutralization of every detectable bad attribute left the mega $94\%$ intact,
because a redundant mesh of moderately-bad values re-closes around each removed
weld (Exps.~8--10).

\paragraph{Two changes of object.} What worked was changing the unit of
analysis. First, from nodes to \emph{topology}: a sampled-betweenness cut of the
exact union graph located the $\sim$$2{,}000$ load-bearing ids whose removal
disconnects the welded communities, $98\%$ of them attribute-clean and so
invisible to any value rule (Exp.~11). Second, from values to \emph{edges}: a
classifier trained on $2.6$M free within-handle GitHub-id labels scored
candidate pairs and pruned the residual given-name homonym fragments the cut had
isolated (Exps.~6, 12). Only with the mega gone was recall safe to attack: the
same classifier, filtering the pairwise expansion of the previously dormant
cross-project shingle layer, recovered in one experiment more recall than the
entire record had spent, and composing it with GitHub's own \texttt{noreply}
account assertions finished the job (Exps.~14, 17), while precision rose.

\paragraph{Transferable lessons.} Four findings generalize beyond this dataset
and motivate the artifacts we ship. (i) \emph{Structural certificates do not
transfer}: a cut computed on one edge set silently fails on an augmented one,
so the production pipeline re-checks the mega-cluster histogram after every new
edge source. (ii) \emph{Constructed labels embed shortcuts}: a gradient-boosted
classifier read the \texttt{users.\allowbreak noreply.\allowbreak github.com}
substring out of the label construction and collapsed on human labels ($0.999$ in-distribution AUC,
$0.56$ transfer), where a linear model on the same feature was benign; the
shipped classifier is therefore logistic, selected by out-of-distribution
transfer, not in-distribution accuracy. (iii) \emph{Bad values split by intent}: privacy
masks and homonym defaults demand the same non-merge action for opposite
reasons, and only the former may never be re-linked, hence the
\emph{bad-by-attribute}/\emph{local} distinction in \texttt{A2clsFull}.
(iv) \emph{Benchmarks are directional}: a precise-but-small human gold and a
recall-only GitHub-scale GT rank maps oppositely, so we grade on both and ship
provenance tags that let consumers choose their own precision/recall operating
point.

% ===========================================================================
\section{Quality and Validation}
\label{sec:quality}
We validate against two independent ground truths read jointly, plus a
direct commit-coverage census. We report \emph{both} error families a partition
admits, because each is invisible to a benchmark that measures only the other.

\paragraph{Splitting vs.\ clumping.} A clustering can fail two ways. It can
\emph{split}: scatter one person's ids across several clusters (a
\emph{fragmentation} error); the splitting rate is $1-\text{recall}$. Or it can
\emph{clump}: merge distinct people into one cluster (an \emph{over-merge} or
conflation error); the clumping rate is $1-\text{precision}$. The two are not
symmetric in how they are reported. Splitting is exposed by any
within-person recall benchmark; clumping is exposed \emph{only} by a benchmark
that contains, and weights, the conflated region. A large literature on author
disambiguation reports near-perfect precision while leaving clumping
effectively unmeasured: by evaluating on curated samples that never reach the
over-merged region, by reporting recall-only ground truths on which an
over-merge costs nothing, or by sampling pairs in a way that under-weights a
single giant cluster (whose false-pair mass grows quadratically in its size).
We therefore report splitting and clumping side by side, and add a
distribution-level clumping diagnostic, the largest cluster, that a pairwise
average can hide.

\begin{figure}[t]
\centering
\includegraphics[width=0.92\linewidth]{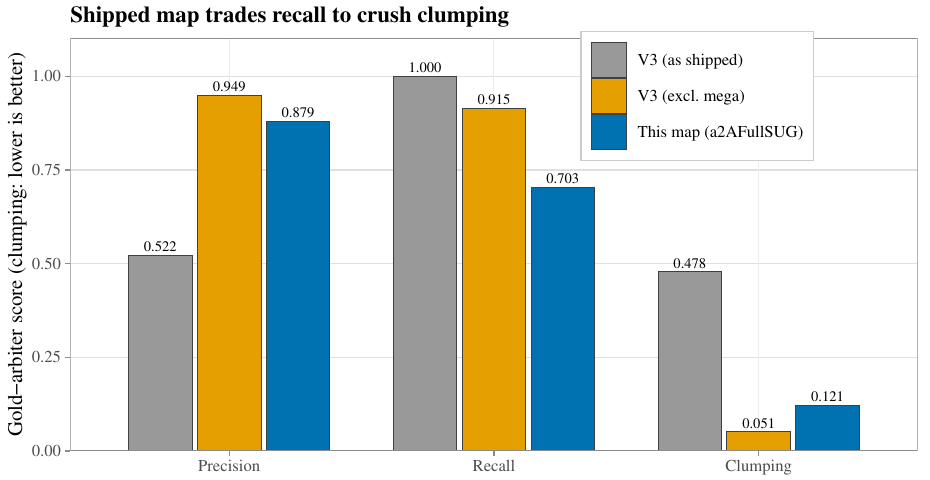}
\caption{Gold-arbiter precision, recall, and clumping (lower is better) for the
shipped \texttt{a2AFullSUG} map against the two V3 framings (Table~\ref{tab:splitclump}):
``as shipped'' counts V3's mega-cluster, ``excl.\ mega'' removes it first.}
\label{fig:splitclump}
\end{figure}

\paragraph{ALFAA human gold (both axes).} The ALFAA pairs~\cite{amreen2019alfaa}
are $469$k human-rated id pairs (label~2 $=$ match), OpenStack-centric.
Table~\ref{tab:splitclump} (visualized in Figure~\ref{fig:splitclump}) scores the
production map against the prior WoC
production map (V3) on this gold. The contrast drives this section. V3
reports \emph{recall $1.000$} (zero splitting); measured after excluding
its over-merged region, the protocol that ``ignores clumping'', it reports
a \emph{mindboggling precision $0.949$}. Measured on the partition \emph{as
shipped}, the same map's precision is $0.522$ and its largest cluster is
$3{,}006{,}318$ ids: $7.7\%$ of all gold ids fall inside that one mega-cluster
and $94.6\%$ of V3's false-positive pairs are produced by it. The
``mindboggling precision'' is an artifact of not counting the clumping. The
production map released here trades a little recall for an honest clumping
profile: splitting $0.297$, clumping $0.121$, largest cluster $6{,}910$ (itself
a single GitHub \texttt{noreply} identity). An audit of its disagreements shows
the gold itself carries $\sim$$26$ mislabels (exact-email pairs rated
``different''), placing its true precision nearer $0.90$.

\begin{table}[t]
\centering\small
\caption{Splitting and clumping on the ALFAA gold. ``V3 (excl.\ mega)'' is the
prior map evaluated with its $3$M-id mega-cluster removed first: the protocol
that reports high precision by not counting clumping. ``V3 (as shipped)'' counts
it. Splitting $=1-$recall; clumping $=1-$precision.}
\label{tab:splitclump}
\begin{tabular}{@{}lrrrr@{}}
\toprule
Map & Prec. & Recall & Clump. & Largest \\
\midrule
V3 (excl.\ mega)   & $0.949$ & $0.915$ & $0.051$ & --- \\
V3 (as shipped)    & $0.522$ & $1.000$ & $0.478$ & $3{,}006{,}318$ \\
\textbf{This map (\texttt{a2AFullSUG})} & $0.879$ & $0.703$ & $0.121$ & $6{,}910$ \\
\bottomrule
\end{tabular}
\end{table}

\paragraph{GitHub ground truth (recall $=$ splitting only).} A complementary,
large-scale recall benchmark derives within-handle alias labels from
single-account GitHub repositories ($9.57$M handles over $21.5$M commit ids; an
independent within-handle ground truth, Bock et al., under
submission~\cite{bock2025dealiasing}). It is the empirical mirror of the
preceding point: it is recall-only, measuring splitting while
\emph{structurally} blind to clumping, because it only contains ids already
known to share one handle, so an over-merge (even a $3$M-id mega) is priced at
zero and the metric even reports the over-merge as a perfect score. Read in
isolation such a benchmark ranks the mega-laden V3 \emph{above} a mega-free map;
read jointly with the clumping axis it is informative. On it the released map
recovers recall $0.60$ under the benchmark's \texttt{max9} scoring and $0.56$
under its stricter \texttt{FullyMerged} scoring, the two recall conventions
that benchmark defines~\cite{bock2025dealiasing}. This is the recall a
\emph{global} map can reach without local repository context; the residual gap
is the within-repository evidence a global map forgoes by design, not
over-merge.

\subsection{Comparison with Prior Approaches}
\label{sec:prior}
The dominant prior VCS name-disambiguation tools are string-similarity merges
validated on small, manually disambiguated corpora, and they are the published
home of the ``mindboggling precision'' this section cautions against. The
email-network heuristic of Bird et al.~\cite{bird2006mining}, the merge
algorithms compared by Goeminne and Mens~\cite{goeminne2013comparison}, and the
LSA-based identity merge of Kouters et al.~\cite{kouters2012whoswho} are evaluated on
single-project or single-community corpora. The most explicit recent example is
the \texttt{gambit} tool of Gote and Zingg~\cite{gote2021gambit}, which reports
an F1 of $0.985$, on a \emph{single} manually disambiguated project (Gnome GTK).
The number is not in dispute; what it cannot reveal is clumping. An evaluation on
a corpus that never contains a global mega-cluster measures splitting while
leaving clumping untested, because the over-merged region that a $10^8$-id union
produces is not in the sample. The same caution applies to the global WoC
maps of the ALFAA lineage~\cite{amreen2019alfaa,fry2020dataset}: as
Table~\ref{tab:splitclump} shows for the ALFAA-lineage map V3, their headline
precision is computed off the conflated region and so does not reveal the
$3{,}006{,}318$-id over-merge sitting in the same partition. We make the opposite
methodological commitment (report splitting and clumping jointly, and ship a
largest-cluster diagnostic), and we contend that any disambiguation result at
scale that reports only one axis is, at present, untestable on the other.

\paragraph{Commit coverage (census).} Computed directly over all
$5{,}866{,}595{,}698$ commits via the \texttt{c2AFull} provenance tags:
$73.5\%$ of all commits resolve into a multi-id identity (up from $66.3\%$ for
the prior baseline map). Restricted to human (non-bot) commits, the global map
alone covers $96.49\%$; adding the within-project layer raises this to
$98.17\%$ ($+1.68$ percentage points), leaving a residual unresolved fraction
of $1.83\%$. Relative to a value-gated union (the baseline before the structural
cut), gold recall rose $0.44\to0.70$
\emph{and} precision rose $0.87\to0.88$, with the largest cluster down from
$170{,}431$ to $6{,}910$.

% ===========================================================================
\section{Downstream Impact: How Aliasing Changes Developer Analytics}
\label{sec:downstream}
Validation (Section~\ref{sec:quality}) characterizes the map's matching
quality: its splitting and clumping rates against two ground truths; this
section asks a different question: \emph{how much does it matter, and to whom?} We recompute a battery of
standard developer-analytics measures at three points on an ablation ladder over the
full World of Code commit corpus ($5{,}866{,}595{,}698$ \texttt{V2604}
commits), so the cost of getting identity wrong is read off
directly rather than argued.

\paragraph{Three method archetypes, two opposite errors.} This section shows
that the \emph{two ways}
of getting identity wrong distort realistic downstream analyses in
\emph{opposite directions}, so a consumer who picks either extreme is misled, in a
predictable direction, on every task. Section~\ref{sec:quality} graded matching
quality (splitting vs.\ clumping); here we trace those two failure families into the
analytics built on top of the map. Three archetypes bracket the design space:
\begin{itemize}
\item \textbf{Under-merge} ($L_0$, raw author string $a=$ name$+$email exact): the
  analyst default of keying on the commit's author string with no resolution at
  all. It \emph{splits} one person across aliases, so it over-counts people and
  under-counts each person's work.
\item \textbf{Over-merge} (naive transitive union over shared-attribute edges; the
  prior WoC map V3 is its shipped instance, with a $3{,}006{,}318$-id
  mega-cluster, Table~\ref{tab:splitclump}). It \emph{clumps} strangers, so it
  under-counts people and welds unrelated work onto a few giant phantom identities.
\item \textbf{Calibrated} ($L_2$, the released map): splitting and clumping both
  bounded, largest cluster $6{,}910$.
\end{itemize}
The under-merge archetype is what our $L_0$ ablation measures directly; the
calibrated archetype is $L_2$; the over-merge archetype is characterized
quantitatively for matching in Table~\ref{tab:splitclump} and its downstream
\emph{direction} follows from its mechanism (collapsing $N$ people onto one node
moves every per-developer aggregate the opposite way from splitting them into $N$).
Table~\ref{tab:archetypes} summarizes the direction of the error each archetype
induces on five common downstream questions; the rest of this section supplies the
measured magnitudes for the two endpoints we run end-to-end ($L_0$ and $L_2$).

\begin{table}[t]
\centering\small
\caption{How each identity-method archetype distorts five realistic downstream
analyses, relative to correct (calibrated) identity. ``$\uparrow$'' $=$
over-states the quantity, ``$\downarrow$'' $=$ under-states it. Under-merge and
over-merge err in \emph{opposite} directions on every row.}
\label{tab:archetypes}
\begin{tabular}{@{}lccc@{}}
\toprule
Downstream question & Under-merge & Over-merge & Calibrated \\
                    & ($L_0$)     & (V3 union) & ($L_2$) \\
\midrule
Developer head-count            & $\uparrow$ $+66.6\%$ & $\downarrow$ & $\approx$ \\
Commits per developer           & $\downarrow$ $-40\%$ & $\uparrow$   & $\approx$ \\
Bus/truck factor (SPOF risk)    & $\uparrow$ (reassuring) & $\downarrow$ (alarmist) & $\approx$ \\
Collaboration-network density   & $\downarrow$ (fragmented) & $\uparrow$ (phantom hub) & $\approx$ \\
Centrality top-ranked devs      & bots/bad ids & the mega-cluster & real devs \\
\bottomrule
\end{tabular}
\end{table}

\begin{table}[t]
\centering
\caption{Ablation ladder. Each level is recovered per commit from the
provenance tag in \texttt{c2AFull} rather than rebuilt as a separate map.}
\label{tab:abllevels}
\begin{tabular}{@{}llp{3.6cm}@{}}
\toprule
Level & Resolution & Ablates \\
\midrule
$L_0$ & raw author string $a$ & all aliasing \\
$L_1$ & core/global map (Exp.~17) & within-project layer \\
$L_2$ & core $+$ within-project (Exp.~18) & nothing (production) \\
\bottomrule
\end{tabular}
\end{table}

\subsection{Developer population and productivity}
\label{sec:abl-pop}
Without aliasing the developer population is \textbf{inflated by $+66.6\%$}:
$104{,}394{,}428$ raw ids collapse to $62{,}670{,}110$ resolved developers (the
$L_0$ ids that \emph{author} commits; fewer than the $106{,}826{,}059$
author/committer strings of Table~\ref{tab:classes}, because committer-only
strings author nothing), so
$40.0\%$ of all raw ids are duplicate identities of someone already counted.
The core map removes the bulk ($-37.8\%$); the within-project layer folds a
further $2{,}245{,}942$ local ids the global map cannot place. The bias is
concentrated, not uniform (Table~\ref{tab:ablmetrics}): the \emph{median}
developer is untouched ($5$ commits and $1$ project under raw ids or production
alike), so the cost of skipping aliasing falls on the \emph{prolific tail},
whose work is split across alias fragments, rather than on typical contributors. At the
$90$th percentile a raw-id study understates output by $34\%$ (p90 commits/dev
$65$ vs.\ $99$) and careers by $34\%$ (p90 span $822$ vs.\ $1{,}243$ days,
$\sim$$14$ months), and it invents $10.1$M one-commit ``developers'' that are
really alias fragments of established contributors. The \emph{median} span moves
only trivially across $L_0$--$L_1$ ($8.1$ to $8.0$ days, a sub-day difference
within rounding for the short-lived bulk of ids) and lifts to $9.3$ only once the
within-project layer stitches a person's repeated local commits into one career;
the career-extension effect of aliasing is a tail phenomenon, visible at p90, not
at the median. (Per-developer means are
omitted by design: with a maximum of $114.7$M commits on a single id, the
distribution is power-law and its mean reflects bot and mega-cluster placement,
not a representative developer.)

\begin{table}[t]
\centering
\caption{Developer metrics under ablation. ``Raw bias'' compares the
no-aliasing $L_0$ view against production $L_2$. Per-developer quantities are
reported as median and p90 rather than the mean, because the distributions are
power-law.}
\label{tab:ablmetrics}
\begin{tabular}{@{}lrrrl@{}}
\toprule
Metric & $L_0$ raw & $L_1$ core & $L_2$ final & Raw bias \\
\midrule
distinct developers   & $104.39$M & $64.92$M & $62.67$M & $+66.6\%$ \\
median commits/dev    & $5$ & $5$ & $5$ & $\approx$ \\
p90 commits/dev       & $65$ & $95$ & $99$ & $-34\%$ \\
median projects/dev   & $1$ & $1$ & $1$ & $\approx$ \\
p90 projects/dev      & $7$ & $9$ & $9$ & $-22\%$ \\
median span (days)    & $8.1$ & $8.0$ & $9.3$ & $-13\%$ \\
p90 span (days)       & $822$ & $1207$ & $1243$ & $-34\%$ \\
1-commit ``devs''     & $25.02$M & $15.82$M & $14.92$M & $+67.7\%$ \\
\bottomrule
\end{tabular}
\end{table}

\subsection{Bad, privacy, and bot identities}
\label{sec:abl-class}
A second hazard is that the bad-attribute, privacy-local, and bot ids the map
neutralizes (the last flagged with the WoC bot detector of
Dey et al.~\cite{dey2020detecting}) would, untreated, be counted as legitimate
developers. They are only
$5.42\%$ of raw ``developers'' but $8.46\%$ of commits (over-represented
$1.56\times$) and they dominate the head of any productivity ranking
(Table~\ref{tab:ablclass}): $23$ of the top $100$ and $169$ of the top $1{,}000$
most-prolific raw ids are illegitimate. The single busiest ``developer'' by
commits is \texttt{dependabot[bot]} ($114.7$M commits over $5.32$M projects);
others masquerading as prolific humans include \texttt{Your Name
<you@example.com>} ($9.1$M), \texttt{Automated} ($6.7$M), and \texttt{GitHub
Action} ($5.9$M). This is the concrete payoff of the bad-attribute, blank, and
bot guards.

\begin{table}[t]
\centering
\caption{Raw-id population stratified by class (no aliasing, $L_0$). Illegitimate
$=$ bad-attribute $+$ local $+$ bot.}
\label{tab:ablclass}
\begin{tabular}{@{}lrrr@{}}
\toprule
Class & Ids & Id\,\% & Commit\,\% \\
\midrule
good            & $98{,}492{,}759$ & $94.35$ & $91.10$ \\
bot             & $541{,}186$      & $0.52$  & $5.13$ \\
bad-attribute   & $2{,}602{,}192$  & $2.49$  & $2.72$ \\
local (privacy) & $2{,}519{,}593$  & $2.41$  & $0.61$ \\
partial         & $238{,}698$      & $0.23$  & $0.45$ \\
\midrule
\textbf{illegitimate} & $\mathbf{5{,}662{,}971}$ & $\mathbf{5.42}$ & $\mathbf{8.46}$ \\
\bottomrule
\end{tabular}
\end{table}

\subsection{The developer collaboration network}
\label{sec:abl-net}
Identity errors miscount individuals, but they also deform the
\emph{structure} of the collaboration graph that an enormous social-network and
mining literature is built on. We construct the co-project developer network at
each ablation level (an undirected edge joins two developers who share at least
one project), capping fan-out at $1{,}000$ distinct developers per project to
exclude the $\sim$$3{,}000$ mega-projects whose $N^2$ cliques are not meaningful
collaboration (the largest raw project alone names $1{,}017{,}654$ ids). The
graph is large: $2.34$B undirected edges at $L_0$.

The raw network systematically \emph{overstates} collaboration
(Table~\ref{tab:abldeg}). Aliasing removes $32\%$ of all edges and cuts the
apparent collaborating population from $70\%$ to $49\%$ of developers, because in
the raw data one person's several ids co-occurring in a project link
``to themselves'': a spurious edge that vanishes once the ids merge. Among the
developers who remain connected, the \emph{typical} node is unchanged (median
degree is $3$ at every level), but the connectivity of the active tail
\emph{rises} (p90 degree $98\to231$) as the low-degree alias fragments that
surround each real person are absorbed into a single, better-connected node. The
most-connected raw node has $10.6$M neighbors; after aliasing the maximum falls
to $6.0$M.

\begin{table}[t]
\centering
\caption{Co-project network under ablation (fan-out cap $1{,}000$).}
\label{tab:abldeg}
\begin{tabular}{@{}lrrr@{}}
\toprule
 & $L_0$ raw & $L_1$ core & $L_2$ final \\
\midrule
nodes (developers)    & $104.39$M & $64.92$M & $62.67$M \\
connected             & $70.0\%$  & $52.3\%$ & $49.0\%$ \\
isolated              & $30.0\%$  & $47.7\%$ & $51.0\%$ \\
unique edges          & $2.34$B   & $1.61$B  & $1.59$B \\
median degree (conn.) & $3$       & $3$      & $3$ \\
p90 degree (conn.)    & $98$      & $196$    & $231$ \\
max degree            & $10.63$M  & $6.24$M  & $6.04$M \\
single-dev projects   & $63.7\%$  & $82.1\%$ & $84.0\%$ \\
\bottomrule
\end{tabular}
\end{table}

The last row is a finding in its own right: $36.3\%$ of projects look
multi-developer in raw ids, but only $16\%$ are after aliasing: roughly half of
all apparent ``collaborations'' are a single person committing under several
identities.

\paragraph{The same pattern on the finer co-file network.}
Sharing a project is a loose notion of collaboration; co-editing the same file
is a stricter one. We rebuild the network with an edge between two developers
only when they both touch the same $(\text{project},\text{file})$ pair (cap
$1{,}000$ co-editors per file), over the same node universe, and run the identical
ablation (Table~\ref{tab:abldegfile}). The co-file graph is sparser, as expected
from the stronger join: $62.3\%$ of developers are connected at $L_0$ against
$70.0\%$ on the co-project graph. Every aliasing effect reappears with the same
sign and nearly the same magnitude. Aliasing again removes $32\%$ of edges
($2.83$B$\to$$1.92$B), drops the connected population from $62.3\%$ to $45.6\%$,
raises the active-tail p90 degree ($96\to231$) as alias fragments coalesce, and
lowers the maximum degree ($5.77$M$\to$$3.50$M). That a structurally distinct,
more conservative collaboration graph reproduces the co-project result rules out
the deformation being an artifact of the project-membership edge rule.

\begin{table}[t]
\centering
\caption{Co-file network under ablation (edge $=$ shared $(\text{project},\text{file})$,
fan-out cap $1{,}000$). Same node universe as Table~\ref{tab:abldeg}.}
\label{tab:abldegfile}
\begin{tabular}{@{}lrrr@{}}
\toprule
 & $L_0$ raw & $L_1$ core & $L_2$ final \\
\midrule
nodes (developers)    & $104.39$M & $64.92$M & $62.67$M \\
connected             & $62.3\%$  & $48.0\%$ & $45.6\%$ \\
isolated              & $37.7\%$  & $52.0\%$ & $54.4\%$ \\
unique edges          & $2.83$B   & $1.95$B  & $1.92$B \\
median degree (conn.) & $3$       & $3$      & $3$ \\
p90 degree (conn.)    & $96$      & $198$    & $231$ \\
max degree            & $5.77$M   & $3.58$M  & $3.50$M \\
\bottomrule
\end{tabular}
\end{table}

\paragraph{Bad ids hide in low-degree space.}
A natural objection is that the deformation is confined to a few obvious hubs
that a degree threshold could delete. It cannot. Table~\ref{tab:abldegcls}
stratifies node degree in the raw network by id class: across every
illegitimate class the hub fraction (degree $\geq 1{,}000$) is under half a
percent ($0.38\%$ of bots, $0.49\%$ of bad-attribute, $0.12\%$ of local ids),
while $70$--$84\%$ sit at degree $\leq 10$ and a large share are fully isolated
($62.5\%$ of bots). The handful of bot \emph{hubs} (the maximum-degree node in
the whole graph is a bot) are what makes ``bots are hubs'' intuitive, yet
the bot \emph{population} is overwhelmingly peripheral. A centrality cutoff would
miss $99.5\%$ of each bad class; separating them requires the attribute- and
name-based classifier, not a structural threshold.

\begin{table}[t]
\centering
\caption{Degree by id class in the raw ($L_0$) network. ``Hub'' $=$ degree
$\geq 1{,}000$.}
\label{tab:abldegcls}
\begin{tabular}{@{}lrrrr@{}}
\toprule
Class & Ids & isol.\,\% & deg$\leq$10\,\% & hub\,\% \\
\midrule
good      & $98.49$M & $30.3$ & $81.9$ & $0.78$ \\
bot       & $541$K   & $62.5$ & $69.9$ & $0.38$ \\
bad-attr  & $2.60$M  & $25.8$ & $78.7$ & $0.49$ \\
local     & $2.52$M  & $15.4$ & $83.9$ & $0.12$ \\
\bottomrule
\end{tabular}
\end{table}

\subsection{Centrality measures}
\label{sec:abl-cent}
Who, structurally, are the most important developers? Centrality on the
collaboration network answers this for influence studies, key-developer detection,
and knowledge-broker analyses. We rank developers in the raw ($L_0$) network by
five standard measures ($k$-core, PageRank, eigenvector, Katz, and approximate
betweenness) and inspect the class composition of the top of each ranking
(Table~\ref{tab:abltopk}). The finding is that on the un-aliased network
\emph{the top of the ranking is mostly not real developers}: of the $50$
highest-betweenness nodes $50\%$ are bots or bad-attribute placeholders, rising to
$60\%$ for PageRank, $64\%$ for eigenvector, and $68\%$ for Katz centrality: the
shared-placeholder ids (\texttt{root}, \texttt{your-name}, CI bots) sit between
otherwise unconnected projects and so look like the graph's most important
brokers. The one exception is $k$-core, whose top $50$ are all legitimate: degeneracy
ordering rewards dense mutual collaboration, which placeholder hubs (high degree but
tree-like, not clique-like) do not have, a useful robustness contrast. This is the
centrality-space face of the same contamination Table~\ref{tab:abldegcls} shows in
degree space: illegitimate ids are not removable by a degree threshold, and on four
of five centrality measures they \emph{dominate} the top ranks an analyst
would read first.

Running the same rankings on the aliased networks ($L_1$, $L_2$) shows that
aliasing \emph{reduces but does not eliminate} the contamination, and reveals
\emph{why} (Table~\ref{tab:abltopk}). Where the residue is bad-attribute
placeholders the merge clears them: approximate betweenness falls from $50\%$ to
$38\%$ non-legitimate and eigenvector from $64\%$ to $48\%$ as the shared
\texttt{your-name}-style ids that bridged unrelated projects coalesce. But the
bulk of the residue is \emph{bots}, and a CI bot is a genuinely distinct
automated account, not a human alias, so the map correctly does not merge it:
the bot count in the betweenness top $50$ falls only from $16$ to $13$, and on
PageRank and Katz, where bots already dominate, the non-legitimate share barely
moves ($60\%\!\to\!58\%$, $68\%\!\to\!62\%$). The lesson reinforces the
degree-space one: the alias map fixes the human-fragmentation half of the
problem, but cleaning a centrality ranking still needs the bot/bad-attribute
\emph{classifier}, not the map alone and not a structural threshold. The headline
holds throughout: centrality on raw identities ranks the noise above the signal.

\paragraph{Centrality on the co-file network.}
Repeating the centrality ablation on the finer co-file graph
(Table~\ref{tab:abltopkfile}) shows the same contamination, generally milder:
co-editing a file is a stronger collaboration signal than sharing a project, so
placeholder ids that merely co-occur in projects rarely co-edit the same file.
Two measures are already robust at $L_0$: $k$-core ($0\%$) and, unlike on the
co-project graph, eigenvector ($2\%$), whose mass concentrates on dense
file-coediting teams rather than placeholder stars. PageRank and Katz are
contaminated in the raw network ($60\%$, $58\%$) and aliasing reduces both (to
$50\%$ and $38\%$ at $L_2$), with the same bot-dominated residue as on the
co-project graph. Approximate betweenness on this $2.83$-billion-edge network
starts at the same $50\%$ contamination at $L_0$ but cleans further than on the
co-project graph as aliasing proceeds, falling to $36\%$ at $L_1$ and $28\%$ at
$L_2$; the raw top-$50$ holds $14$ bots and $11$ bad-attribute placeholders, and
the map clears the placeholders while the bots, not an aliasing target, persist.

\begin{table}[t]
\centering\small
\caption{Illegitimate-id share of the top-$50$ developers on the \emph{co-file}
network across the ablation ladder (cf.\ Table~\ref{tab:abltopk} for co-project).
Contamination is milder than on the co-project graph and eigenvector is robust
here.}
\label{tab:abltopkfile}
\begin{tabular}{@{}lrrrl@{}}
\toprule
Centrality measure & $L_0$ & $L_1$ & $L_2$ & $L_0$ make-up \\
\midrule
$k$-core (degeneracy)    & $0\%$  & $4\%$  & $0\%$  & $50$ good \\
Eigenvector              & $2\%$  & $4\%$  & $4\%$  & $49$ good, $1$ bad-attr \\
PageRank                 & $60\%$ & $50\%$ & $50\%$ & $20$ good, $20$ bot, $10$ bad-attr \\
Katz                     & $58\%$ & $38\%$ & $38\%$ & $21$ good, $23$ bot, $6$ bad-attr \\
Approx.\ betweenness     & $50\%$ & $36\%$ & $28\%$ & $25$ good, $14$ bot, $11$ bad-attr \\
\bottomrule
\end{tabular}
\end{table}

\begin{table}[t]
\centering\small
\caption{Illegitimate-id share of the $50$ top-ranked developers on the
co-project network under five centrality measures, across the ablation ladder.
``Non-legit.\,\%'' counts bot and bad-attribute ids among the top $50$; the last
column gives the raw ($L_0$) make-up.}
\label{tab:abltopk}
\begin{tabular}{@{}lrrrl@{}}
\toprule
Centrality measure & $L_0$ & $L_1$ & $L_2$ & $L_0$ make-up \\
\midrule
$k$-core (degeneracy)    & $0\%$  & $2\%$  & $0\%$  & $50$ good \\
Approx.\ betweenness     & $50\%$ & $42\%$ & $38\%$ & $25$ good, $16$ bot, $9$ bad-attr \\
PageRank                 & $60\%$ & $62\%$ & $58\%$ & $20$ good, $21$ bot, $9$ bad-attr \\
Eigenvector              & $64\%$ & $48\%$ & $48\%$ & $18$ good, $28$ bot, $4$ bad-attr \\
Katz                     & $68\%$ & $62\%$ & $62\%$ & $16$ good, $26$ bot, $8$ bad-attr \\
\bottomrule
\end{tabular}
\end{table}

\subsection{Team resilience: bus/truck factor, reviewers, and successors}
\label{sec:abl-bus}
Three of the most common operational questions a project-analytics system answers
are: who are the few people the project cannot afford to lose (\emph{bus} or
\emph{truck factor}~\cite{avelino2016truck}), who is qualified to review a change (\emph{reviewer
recommendation}), and who could take over a departing contributor's work
(\emph{successor recommendation}). All three are computed from the same
substrate (how commits concentrate on developers) and all three are distorted
in the same direction by identity fragmentation.

We compute a commit-concentration bus factor for all $189.9$M projects at each
ablation level: the minimum number of developers whose cumulative share first
reaches $50\%$ of the project's commits. Under-aliased ids ($L_0$) split one
person's work across several identities, so each holds fewer commits and the
project appears to spread its knowledge over \emph{more} people than it really
does: an inflated, falsely reassuring bus factor. Because merging identities can
only concentrate commits onto fewer developers, the bus factor is
\emph{monotone non-increasing} along the ablation ladder: every change is a
correction in the same direction (Table~\ref{tab:ablbus}).

\begin{table}[t]
\centering
\caption{Bus/truck factor under ablation. \textsc{bf}$=$min developers covering
$50\%$ of a project's commits; \emph{solo}$=$projects with one distinct
developer; \emph{SPOF}$=$single point of failure (\textsc{bf}$=1$).}
\label{tab:ablbus}
\begin{tabular}{@{}lrrr@{}}
\toprule
 & $L_0$ raw & $L_1$ core & $L_2$ final \\
\midrule
projects             & $189.86$M & $189.86$M & $189.86$M \\
SPOF (\textsc{bf}$=1$) & $90.15\%$ & $96.05\%$ & $96.48\%$ \\
solo (1 developer)   & $63.71\%$ & $82.15\%$ & $83.97\%$ \\
projects with \textsc{bf}$\geq2$ & $18.71$M & $7.49$M & $6.68$M \\
\bottomrule
\end{tabular}
\end{table}

The correction is large and one-directional. The share of projects resting on a
\emph{single} indispensable developer rises from $90.15\%$ to $96.48\%$, and the
count of projects that appear to have a bus factor of two or more falls from
$18.71$M to $6.68$M. Equivalently, $12.0$M projects ($64.3\%$ of every project
the raw data rates as collaborative) collapse to a single point of failure once
aliases are merged: projects an analyst would wrongly judge ``safe.'' Most of the
correction comes from the global core map ($L_0\!\to\!L_1$, solo $63.7\%\!\to\!82.2\%$),
with within-project matching adding the rest ($L_1\!\to\!L_2$, $\to\!84.0\%$). The
solo-project share, reached here from commit concentration, independently matches
the $84.0\%$ obtained from the collaboration graph (\S\ref{sec:abl-net}). Two
different computations converge on the same fact: roughly a fifth of all
projects that look multi-developer are one person under several identities.

\paragraph{File-level resilience.}
The project-level bus factor is a coarse instrument: reviewer and successor
recommendation operate at the granularity of the individual \emph{file}, not the
whole project. We therefore recompute the same concentration measure over every
$(\text{project},\text{file})$ pair in WoC ($7.49$ billion units) at each
ablation level (Table~\ref{tab:ablbusfile}). The pattern is identical in
direction and, because files have far fewer touching developers than projects,
even more sharply concentrated: the share of files edited by a single developer
rises from $85.06\%$ ($L_0$) to $90.58\%$ ($L_2$), and the apparent
multi-developer share is nearly halved, $14.94\%\!\to\!9.42\%$. As at the project
level the correction is overwhelmingly the global core map ($L_0\!\to\!L_1$), with
within-project matching trimming the residue.

\begin{table}[t]
\centering
\caption{Bus/truck factor at the \emph{file} granularity, over all
$7.49$B $(\text{project},\text{file})$ pairs. \textsc{solo}$=$files touched by one
distinct developer; \emph{multi}$=$files that appear to have $\geq2$.}
\label{tab:ablbusfile}
\begin{tabular}{@{}lrrr@{}}
\toprule
 & $L_0$ raw & $L_1$ core & $L_2$ final \\
\midrule
\textsc{spof} (\textsc{bf}$=1$) & $91.14\%$ & $94.84\%$ & $95.07\%$ \\
solo (1 developer)    & $85.06\%$ & $90.19\%$ & $90.58\%$ \\
apparent multi-dev    & $14.94\%$ & $9.81\%$  & $9.42\%$ \\
\bottomrule
\end{tabular}
\end{table}

\paragraph{Reviewers and successors.}
The same fragmentation that inflates the bus factor degrades reviewer and
successor recommendation, because both rank candidates by accumulated experience
on the relevant files or project. When a contributor's history is scattered
across aliases, three failures follow. (i) A genuine expert can be \emph{passed
over}: each of their aliases looks like a newcomer below the experience
threshold, so a less-qualified ``whole'' id is recommended instead. (ii) A
recommended reviewer and the author can be the \emph{same person} under two
emails, a self-review the system cannot detect. (iii) For successor
recommendation the error is sharpest: a developer who merely changed their email
looks like they \emph{left} while a stranger \emph{arrived}, firing a spurious
succession event and proposing a successor for work whose author never departed.

The file-level ablation lets us put a number on each of these failures rather than
merely assert them. (i) Of the $1.12$ billion files that \emph{appear}
multi-author in the raw data ($\text{ndev}_{L_0}\!>\!1$), $36.95\%$
($413.6$M files) resolve to a \emph{single} person once aliases are merged: a
``whole'' expert who, before aliasing, presents as several sub-threshold
newcomers. (ii) Of the $4.56$ billion adjacent author hand-offs in file
histories (consecutive commits attributed to different raw ids), $28.10\%$
($1.28$B) are \emph{phantom self-reviews} (the same canonical developer under two
emails), and only $71.90\%$ ($3.28$B) are real hand-offs that survive aliasing.
(iii) $307$M files ($4.10\%$) fire a \emph{phantom succession event}: the
apparent last author before the project's most recent contributor is, after
aliasing, that very contributor, so the successor model would propose replacing a
developer who never left. These are not corner cases; they are the dominant
failure mode of file-history analytics run on raw identities. Consistent with the
within-project matching that produces them
(\S\ref{sec:construction}), which alone reattributes $93.4$M commits ($47.8\%$ of
the low-quality pool) to a developer already active in the same project, correct
identity is a precondition, not a refinement, for any of these tasks.

\subsection{Threats to validity (downstream comparison)}
\label{sec:abl-threats}
The directional claim of Table~\ref{tab:archetypes} (that under-merge and
over-merge bias every analysis in opposite directions) rests on two kinds of
evidence with different strength, and we are explicit about which is which.
\emph{Construct.} The two \emph{endpoints} we run end-to-end are $L_0$ (under-merge)
and $L_2$ (calibrated); their magnitudes are measured over the full corpus. The
calibrated reference $L_2$ is itself imperfect (at recall $0.70$ it still leaves
some of one person's aliases split), so every under-merge bias reported against it
is a conservative \emph{lower} bound: the gap to error-free identity is at least as
large as shown, never smaller (a still-split $L_2$ over-counts developers, so the
true head-count inflation exceeds the measured $+66.6\%$). The
over-merge column is \emph{directional}, not a re-run of V3 through the same
analytics: its sign follows deductively from its mechanism (collapsing $N$ distinct
people onto one identity moves each per-developer aggregate monotonically opposite
to splitting one person into $N$ aliases), and its one anchored magnitude is the
$3{,}006{,}318$-id mega-cluster of Table~\ref{tab:splitclump}. Materializing the
numeric over-merge column by pushing V3 through this same ablation is the natural
next step and would convert each ``$\uparrow$/$\downarrow$'' into a number.
\emph{Internal.} $L_1$ is recovered per commit from the provenance tag rather than
rebuilt as a separate map, so a commit resolved only by within-project matching
reverts to its raw id under $L_1$ (Table~\ref{tab:abllevels}); this is exact, not
an approximation, but it means $L_1$ is a \emph{lower} bound on what a core-only map
recovers. \emph{External.} The collaboration network uses a fan-out cap of $1{,}000$
co-developers per project (Table~\ref{tab:abldeg}), which truncates the few
mega-projects and so \emph{under}-states, not over-states, the density gap; the
centrality ablation is complete across $L_0$/$L_1$/$L_2$ on both the co-project
(Table~\ref{tab:abltopk}) and co-file (Table~\ref{tab:abltopkfile}) networks. None
of these threatens the
\emph{sign} of the comparison, which is the paper's claim; they bound its
\emph{magnitude}.

% ===========================================================================
\section{Broader Impact: Linking Software Authors to Scholarship}
\label{sec:impact}
A canonical software-author identity is useful inside software engineering, but
its larger promise is as a join key \emph{across} corpora. The people who write
code also write papers, datasets, and documentation; tying a developer's
WoC-scale software identity to their scholarly record would let us trace
research-software provenance, measure the scientific impact of code, credit
research-software engineers, and study how software and the literature that
describes it co-evolve.

Two large scholarly author graphs make this concrete. OpenAlex and the
Semantic Scholar author graph each enumerate on the order of $10^8$ authors with
names, affiliations, external identifiers (ORCID, DOI/DBLP linkages), and
publication/citation counts. A local extract of the Semantic Scholar graph holds
$111{,}255{,}251$ authors (\texttt{authorid}, name, aliases, affiliations,
paper/citation counts, $h$-index), the same order of magnitude as the WoC
identity space released here ($1.07\times10^8$ raw ids, folding to $62.67$M
canonical developers). Linking the two is an entity-resolution problem at the
$10^8$ scale~\cite{christen2012matching}, which aliasing first reduces on the WoC
side to those $62.67$M canonical identities --- matching the raw strings directly
would reintroduce exactly the clumping the map exists to prevent.

\paragraph{Clumping, again, is the binding constraint.} A prototype match of WoC
author strings to OpenAlex authors by name and email produces $20{,}658{,}243$
candidate links, of which only $14.6\%$ carry an ORCID and $41.6\%$ point to a
scholarly author with a \emph{single} recorded work: the signature of a
spurious match to a near-empty profile. Single-token or generic names
(\texttt{P}, \texttt{00}, bare given names) match scholarly homonyms wholesale.
The Semantic Scholar side is no richer: of its $111{,}255{,}251$ authors,
$99.8\%$ carry no affiliation, $95.8\%$ no ORCID/DBLP external id, and $50.1\%$ a
single recorded work, so attribute matching there collapses to name alone, and
names are catastrophically ambiguous, with $57{,}850{,}738$ distinct normalized
names, one fifth ($20.8\%$) shared by two or more authors, and the most common
name held by $148{,}190$ distinct authors.
This is the clumping failure of Section~\ref{sec:quality}, now across
corpora and with the multiplicative blow-up of two large namespaces: a name
shared by $k$ developers and $m$ researchers admits $k\times m$ false links, and
a single careless transitive step fuses an entire WoC cluster into one scholar,
or one scholar into a WoC mega-cluster. Cross-corpus linking is thus the most
precision-hostile setting we have encountered, and a recall-oriented evaluation
would again hide the damage.

\paragraph{Why the released map is the right substrate.} Two properties of this
release make it the prerequisite for honest cross-corpus linking. First,
mega-cluster freedom: links flow \emph{through} WoC identities, so a WoC
over-merge propagates directly into the scholarly graph; linking through a
$3$M-id mega would attach three million people to whatever paper any one of them
coincidentally name-matches. Second, the per-id quality classification and
provenance tags let a linker restrict to the high-confidence \emph{good} ids
that carry a real name and email (the only ids on which a match to a scholarly
author is even meaningful) and weight down or discard the bad-by-attribute and
local ids that generate most spurious links.

\paragraph{A DOI-anchored pilot.} To show that the released map makes honest
cross-corpus linking tractable, not merely possible, we ran a pilot that
\emph{anchors} each candidate link on independent corroboration rather than name
agreement alone, the same ``corroboration de-clumps'' principle that drives the
within-corpus construction. Many repositories cite the literature they implement:
we extracted DOIs from \texttt{README}/\texttt{CITATION} blobs, yielding
$1{,}322{,}124$ distinct DOIs across $164{,}655$ projects, of which $53.0\%$
resolve to a paper in the Semantic Scholar crosswalk
($\mathrm{DOI}\to\mathrm{paper}\to\{\mathrm{authorId},\mathrm{name}\}$,
$4.6\times10^8$ rows). We key on the \emph{canonical} author \texttt{A}, building
an idealized record (name variants and organizational email domains) from the
union of all of \texttt{A}'s raw member ids, and accept a link only when a
project's WoC author and a Semantic Scholar author of a DOI \emph{that project
cites} agree on name. This yields $7{,}626$ confident links over $7{,}246$
canonical WoC authors ($74.6\%$ of which carry an organizational email) and
$7{,}574$ scholarly authors.

The anchoring is what makes the result trustworthy. The matched authors
would, under name-only matching, each map to an average of $63.8$
Semantic Scholar authors (the most ambiguous to $35{,}835$); requiring a shared
DOI collapses this to $1.05$ scholarly authors per WoC author. The resulting
bipartite graph has no mega-cluster (its largest connected component spans $25$
nodes) and is almost perfectly $1{:}1$: only $301$ WoC authors link to two or
more scholarly ids (candidates for splits \emph{within} Semantic Scholar) and
only $51$ scholarly ids to two or more WoC authors (residual homonyms). For the
$301$ candidate within-corpus splits, paper-level SPECTER\,v2
embeddings~\cite{cohan2020specter}
corroborate the merge: the mean pairwise cosine among an author's linked papers
is $0.884$ versus a $0.808$ random-pair baseline, an independent recall lever
that cannot re-clump because it fires only inside an already DOI-anchored group.
The cost of this precision is coverage ($7{,}246$ authors is small against the
WoC scale, bounded by the $53\%$ DOI-in-corpus rate and the in-project
name-agreement requirement), the trade a clumping-graded
evaluation is meant to expose. We release the map as the substrate for this
linking; the pilot establishes that DOI co-occurrence (and, inside its anchored
groups, embeddings and organizational email) is a precise, scalable join, and
scaling it (graded explicitly for clumping) is ongoing work.

% ===========================================================================
\section{Usage and Research Opportunities}
\label{sec:usage}
\paragraph{Basic use.} To attribute a commit, look it up in \texttt{c2AFull}
for its resolved author \texttt{A} and provenance tag; or, starting from a raw
author string, look it up in \texttt{a2AFullSUG} for its canonical \texttt{A}.
A study chooses a quality bar by filtering on the provenance tag or the
\texttt{A2clsFull} class, e.g.\ ``human, globally resolved'' is
\texttt{prov$\in\{g,p\}$} and \texttt{class}$\neq$\texttt{bot}.

\paragraph{Opportunities.} The release enables (i) bias-corrected
contribution and bus-factor studies that no longer inherit a $3$M-id
over-merge; (ii) developer-network and migration analyses at full-WoC scale
with an explicit, per-commit resolution-quality signal; (iii) study of the
remaining hard cases (same person, disjoint rare emails), which are the natural
territory of behavioral fingerprinting~\cite{amreen2019alfaa} layered over this
map; and (iv) reuse of the \emph{construction} as a template: the
gate-then-cut-then-classify ordering, and the two-benchmark grading, transfer
to identity resolution in other large heterogeneous graphs.

% ===========================================================================
\section{Limitations}
\label{sec:limits}
The map is global and so omits within-repository evidence that a
repository-local resolver can use; the residual $1.83\%$ unresolved human
commits and the ``same person, disjoint rare emails'' misses are its known
recall frontier. Quality classes are heuristic (name-spread plus a
frequency/blacklist battery) and inherit the usual false-positive/false-negative
trade of attribute heuristics. The map is a snapshot of WoC \texttt{V2604};
identities accrue new aliases over time, so the map is re-derived each WoC
release. Finally, the GitHub ground truth covers only within-handle pairs and
the ALFAA gold is OpenStack-centric, so neither benchmark is a complete oracle,
which is why we report both and ship per-commit provenance rather than
a single quality verdict.

% ===========================================================================
\section{Conclusion}
\label{sec:concl}
We release a mega-cluster-free author-identity map for the World of Code
\texttt{V2604}, resolving $106{,}826{,}059$ raw strings over
$5{,}866{,}595{,}698$ commits, together with a per-id quality classification, a
within-project fallback, and per-commit provenance. The design is an \emph{ordering}
rather than a single algorithm (gate, then cut topology, then classify edges,
then recover recall, then resolve locally), each step forced by a measured
failure of the simpler alternative. The result resolves $73.5\%$ of all six
billion commits and $98.17\%$ of human commits into a multi-id identity, at
recall $0.70$ / precision $0.88$ on human gold. The choice of map is not
cosmetic: left unaliased, the same corpus inflates the developer head-count by
$66.6\%$ and skews productivity, team-resilience, and centrality measures, with
under-merge and over-merge distorting them in opposite directions. The artifacts
and a self-contained replication package ship with the WoC \texttt{V2604} release.

\paragraph{Future work.} The map is the substrate for cross-corpus author
linking (Section~\ref{sec:impact}): resolving these $62.67$M canonical software
identities (folded from $10^8$ raw ids)
against the $10^8$-scale OpenAlex and Semantic Scholar author graphs, so that
software artifacts can be tied to their academic provenance and vice versa. A
DOI-anchored pilot already links $7{,}246$ canonical authors at $1.05$ scholarly
ids each (mega-cluster-free where name-only matching would average $63.8$), so
the open problem is scaling coverage beyond the DOI-cited subset without
surrendering that precision. Because this is the most clumping-hostile setting we
have met, the contribution there will be as much the \emph{evaluation
discipline} (grading the linkage for over-merge, not recall) as the linkage
itself.

\bibliographystyle{ACM-Reference-Format}
\bibliography{aliasing}
\end{document}